\documentclass{article}
\title{Escort density operators and\\
generalized quantum information measures}
\author{Jan Naudts\\
\small Departement Fysica, Universiteit Antwerpen,\\
\small Universiteitsplein 1, 2610 Antwerpen, Belgium\\
\small E-mail Jan.Naudts@ua.ac.be}

\usepackage{amsfonts}
\newcommand{\Ro}{{\mathbb R}}

\newcommand{\Tr}{{\rm  Tr\, }}
\newenvironment{theorem}[1]
{\paragraph {#1}\,\sl}

\newenvironment{proof}
{\par\vskip 8pt\noindent {\bf Proof}}
{\strut\hfill$\square$\par\vskip 0.5cm}

\begin {document}

\maketitle

\begin{abstract}
Parametrized families of density operators are studied. A generalization
of the lower bound of Cram\'er and Rao is formulated. It involves
escort density operators. The notion of $\phi$-exponential family is
introduced. This family, together with its escort, optimizes the
generalized lower bound. It also satisfies a maximum entropy principle
and exhibits a thermodynamic structure in which entropy and free
energy are related by Legendre transform.
\end{abstract}


\section{Introduction}

The exponential family of probability distributions plays a central role in
statistical mechanics, where it is called the Boltzmann-Gibbs distribution.
In a previous paper \cite {NJ05} the author has proposed a generalization
of this family, involving the notion of deformed exponential functions.
It is called the $\phi$-exponential family because it involves a function
$\phi$, which is used to deform the exponential function.
The present paper extends this work to the quantum-mechanical context,
solving some of the difficulties resulting from working with
non-commuting operators.
Other difficulties that can arise (unbounded operators, operators that are not
trace-class, ...) are not discussed at all. However, none of the
latter technicalities arise if the Hilbert space $\cal H$ of
quantum-mechanical wavefunctions is assumed to be finite dimensional.

The motivation for this work is twofold. At one hand the canonical ensemble,
although dominant, is not the only ensemble that occurs in statistical physics.
Hence the present generalization might have practical applications. In fact, in
nonextensive thermostatistics \cite {TC88} , which involves the
$\phi$-exponential family with $\phi(u)=u^q$,
many such applications are claimed. But because of the generality
of the present approach applications
in statistics and in information theory are possible as well.
At the other hand the formulation of a generalization might provide new insights
into the existing theory. Of particular interest is a better understanding
of the problems around the definition of a quantum information manifold
--- see \cite {SRF03} for a recent attempt in this direction. Further work
is needed both on the fundamental level and in the direction of finding
applications.

The paper is organised as follows. The next section generalizes the well-known
lower bound of Cram\'er and Rao in quantum context. In Section 3 a sufficient
condition for optimality of this lower bound is presented. Section 4 introduces
the $\phi$-exponential family of density operators. In Section 5 appropriate
notions of entropy and of relative entropy are defined. Section 6 contains the
obvious example of powerlaw distributions. In Section 7 the $\phi$-exponential
family is shown to satisfy a maximum entropy principle. The associated
structure of Legendre transforms is revealed.
The final section contains a short discussion of results.

\section{Escort density operators and the lower bound of Cram\'er and Rao}

Fix a Hilbert space $\cal H$. Recall that a density operator $\rho$ of
$\cal H$ is a positive operator which is trace-class, with
trace equal to 1.
Let us consider families of density operators $(\rho_\theta)_{\theta\in D}$,
with parameters $\theta=(\theta^1,\cdots,\theta^N)$ in some open domain $D$,
subset of $\Ro^N$.
It is assumed that $\rho_\theta$, seen as a function
of its parameters $\theta$, is sufficiently smooth so that its derivatives
$\partial\rho_\theta/\partial\theta^k$ are well-defined operators.

Consider two families of density operators $(\rho_\theta)_{\theta\in D}$
and $(\sigma_\theta)_{\theta\in D}$ with common parameter domain $D$.
The two families are not related any further. Still, $\sigma_\theta$ is called
the {\sl escort density operator} of $\rho_\theta$.
Averages with respect to the two families of density operators are denoted by
\begin{eqnarray}
\langle A\rangle_\theta=\Tr\rho_\theta A
\qquad\hbox{ resp. }\quad
[A]_\theta=\Tr\sigma_\theta A.
\end{eqnarray}
For simplicity let us assume that $\sigma_\theta$ is invertible.
A generalized Fisher information $g(\theta)$ is defined by
\begin{eqnarray}
g_{kl}(\theta)=
\Tr \frac 1{\sigma_\theta}
\frac{\partial\rho_\theta}{\partial\theta^k}
\frac{\partial\rho_\theta}{\partial\theta^l}
=\bigg[\bigg(\frac 1{\sigma_\theta}
\frac{\partial\rho_\theta}{\partial\theta^k}\bigg)
\bigg(\frac 1{\sigma_\theta}
\frac{\partial\rho_\theta}{\partial\theta^l}\bigg)^\dagger
\bigg]_\theta.
\end{eqnarray}
The $N$-by-$N$-matrix may contain complex entries. However, it is
positive-definite.

Let be given bounded self-adjoint operators $H_1,\cdots H_N$ and assume that a
function $F(\theta)$ exists for which
\begin{eqnarray}
\langle H_k\rangle_\theta
=\frac {\partial F}{\partial\theta^k},\qquad k=1,\cdots N.
\label {aven}
\end{eqnarray}
In physical applications these operators $H_k$ are the macroscopic observables
like Hamiltonian, total magnetization, particle number, ... .
The parameters $\theta_k$ could be inverse temperature, external magnetic
field, chemical potential (multipied with inverse temperature), ... .
The function $F$ is minus the Massieu function (free energy multiplied with inverse
temperature).
The well-known lower bound of Cram\'er and Rao can now be generalized as follows

\begin {theorem}{Proposition}
For arbitrary complex numbers $u^k$ and $v^m$ is
\begin{eqnarray}
\frac {u^k\overline u^l
\left([H_kH_l]_\theta-[H_k]_\theta\,[H_l]_\theta\right)}
{\big|\overline u^mv^n\frac {\partial^2 F}{\partial\theta^m\partial\theta^n}
\big|^2}
\ge
\frac 1{\overline v^mv^ng_{mn}(\theta)}.
\label {genCR}
\end{eqnarray}

\end {theorem}

\begin {proof}
The proof is based on Schwarz's inequality. Let
\begin {eqnarray}
X_k=\frac 1{\sigma_\theta} \frac{\partial\rho}{\partial\theta^k}
\qquad\hbox{ and }\quad
Y_k=H_k-[H_k]_\theta.
\end {eqnarray}
Then one has
\begin{eqnarray}
\bigg|\overline u^kv^m[Y_kX_m^\dagger]_\theta\bigg|^2
&\le&
u^k\overline u^l[Y_kY_l]_\theta
\,
\overline v^nv^m [X_n X_m^\dagger]_\theta.
\label {si}
\end{eqnarray}
The l.h.s.~evaluates to
\begin{eqnarray}
\bigg|\overline u^kv^m[Y_kX_m^\dagger]_\theta\bigg|^2
&=&\bigg|\overline u^kv^m\left[(H_k-[H_k]_\theta)
\frac{\partial\rho}{\partial\theta^m}\frac 1{\sigma_\theta}
\right]_\theta\bigg|^2\cr
&=&\bigg|\overline u^kv^m\Tr (H_k-[H_k]_\theta)
\frac{\partial\rho}{\partial\theta^m}\bigg|^2\cr
&=&\bigg|\overline u^kv^m\frac{\partial\,}{\partial\theta^m}
\langle H_k\rangle_\theta \bigg|^2\cr
&=&\bigg|\overline u^kv^m\frac {\partial^2 F}
{\partial\theta^m\partial\theta^k}\bigg|^2
\end{eqnarray}
Evaluation of the r.h.s.~of (\ref {si}) is straightforward.
The result can be written as (\ref {genCR}).

\end {proof}

\section{Optimal escort families}

The lower bound (\ref {genCR}) is said to be optimal if equality holds
whenever $u^k=v^k$ for all $k$. The exponential family is
known to optimize the usual version of the lower bound of
Cram\'er and Rao. escort density operators are allowed in the
generalized lower bound (\ref{genCR}). One can therefore expect that
other families than the exponential one can optimize the lower bound
provided that the escort family is chosen in a suitable way.
Explicit examples of such families are discussed in the next section.
Here, a sufficient condition for optimizing (\ref{genCR}) is derived.

\begin {theorem} {Proposition}
A sufficient condition under which the escort
family $(\sigma_\theta)_{\theta\in D}$ optimizes (\ref {genCR})
is that there exists functions $G(\theta)$ and $Z(\theta)$
such that
\begin{eqnarray}
\sigma_\theta^{-1}\frac {\partial\rho_\theta}{\partial\theta^k}
=\frac {\partial\rho_\theta}{\partial\theta^k}\sigma_\theta^{-1}
=Z(\theta)\frac {\partial\,}{\partial\theta^k}
\bigg(G(\theta)-\theta^l H_l\bigg).
\label {optcond}
\end{eqnarray}
\end {theorem}

\begin {proof}
In order to show this point first notice that
\begin{eqnarray}
\Tr\sigma_\theta\bigg(\sigma_\theta^{-1}
\frac {\partial\rho_\theta}{\partial\theta^k}\bigg)
&=&\frac {\partial\,}{\partial\theta^k}\Tr\rho_\theta
=0.
\end{eqnarray}
Hence, (\ref {optcond}) implies that
\begin{eqnarray}
\frac {\partial G}{\partial\theta^k}=[H_k]_\theta.
\end{eqnarray}
One then calculates
\begin{eqnarray}
g_{kl}(\theta)
&=&\Tr \frac 1{\sigma_\theta}
\frac{\partial\rho}{\partial\theta^k}
\frac{\partial\rho}{\partial\theta^l}\cr
&=&Z(\theta)^2\Tr\sigma_\theta\bigg([H_k]_\theta-H_k\bigg)
\bigg([H_l]_\theta-H_l\bigg)\cr
&=&Z(\theta)^2\bigg([H_kH_l]_\theta-[H_k]_\theta [H_l]_\theta\bigg).
\label {expg}
\end{eqnarray}
On the other hand is, using (\ref {optcond}),
\begin{eqnarray}
\frac {\partial^2 F}{\partial\theta^k\partial\theta^l}
&=&\frac{\partial\,}{\partial\theta^k}\langle H_l\rangle\cr
&=&\Tr\frac{\partial\rho}{\partial\theta^k}H_l\cr
&=&Z(\theta)\Tr\sigma_\theta
\bigg(\frac {\partial G}{\partial\theta^k}-H_k\bigg)H_l\cr
&=&Z(\theta)\bigg([H_k]_\theta[H_l]_\theta-[H_kH_l]_\theta\bigg).
\label {i2}
\end{eqnarray}
Combining (\ref {expg},\ref {i2}) gives
\begin{eqnarray}
\frac {u^k\overline u^l\left([H_kH_l]_\theta-[H_k]_\theta\,[H_l]_\theta\right)}
{\bigg|\overline u^mu^n\frac {\partial^2 F}{\partial\theta^m\partial\theta^n}
\bigg|^2}
=
\frac 1{\overline u^mu^ng_{mn}(\theta)}.
\end{eqnarray}

\end {proof}

\section{$\phi$-exponential family}

Fix a positive nondecreasing function $\phi(u)$, defined for $u\ge 0$.
Use it to define another function, denoted $\exp_\phi(u)$, by the
relations
\begin{eqnarray}
\frac {{\rm d}\,}{{\rm d}u}\exp_\phi(u)=\phi\bigg(\exp_\phi(u)\bigg)
\qquad\hbox{ and }\quad \exp_\phi(0)=1.
\end{eqnarray}
It is called a deformed exponential \cite {NJ02}.
Its inverse $\ln_\phi(u)$ is a deformed logarithm.

As before, a set of Hamiltonians $H_1,\cdots,H_N$ is fixed.
The $\phi$-exponential family is defined in terms of these Hamiltonians by
\begin{eqnarray}
\rho_\theta=\exp_\phi(G(\theta)-\theta^kH_k).
\end{eqnarray}
It is assumed that these operators $\rho_\theta$ are trace-class.
Normalization $\Tr\rho_\theta=1$ determines the function $G(\theta)$.

Assume in what follows that the operators $H_1,\cdots, H_N$ are two-by-two
commuting. This condition is needed for easy calculation of the derivatives
$\partial\rho_\theta/\partial\theta^k$. With this assumption there exists
a common orthonormal basis of eigenvectors $(\psi_n)_n$
\begin{eqnarray}
H_k\psi_n=\epsilon_{k}(n)\psi_n.
\end{eqnarray}
Then also $\rho_\theta$ is diagonal in the same basis and one has
\begin{eqnarray}
\rho_\theta\psi_n=\exp_\phi\bigg(G(\theta)-\theta^k\epsilon_{k}(n)\bigg)\psi_n.
\end{eqnarray}
and
\begin{eqnarray}
\frac {\partial\rho_\theta}{\partial\theta^k}\psi_n
&=&\phi\bigg(\exp_\phi\bigg(G(\theta)-\theta^k\epsilon_{k}(n)\bigg)\bigg)
\bigg(\frac {\partial G}{\partial\theta^k}-\epsilon_k(n)\bigg)\psi_n\cr
&=&\phi(\rho_\theta)\bigg(\frac {\partial G}{\partial\theta^k}-H_k\bigg)\psi_n.
\end{eqnarray}
Using this last result it is now straightforward to verify that the
condition for optimality (\ref {optcond})
is satisfied with escort family and normalization given by
\begin{eqnarray}
 \sigma_\theta=\frac 1{Z(\theta)}\phi(\rho_\theta)
 \qquad\hbox{ and }\quad
Z(\theta)=\Tr\phi(\rho_\theta).
\end{eqnarray}

\section {Entropy and relative entropy functionals}

The $\phi$-exponential family is the basis for formulating a
generalized thermostatistics \cite {NJ03b,NJ04}. The derivation of the
thermodynamical structure requires the introduction of an entropy functional
$I_\phi(\rho)$ and of relative entropy $I_\phi(\rho||\sigma)$.
The classical version of these quantities is found in \cite {NJ05}, although
the way of presenting here is slightly different.

An appropriate definition of entropy is
\begin{eqnarray}
I_\phi(\rho)
&=&-\int_0^1{\rm d}x\,\Tr\rho\ln_\phi(x\rho)+\int_0^1{\rm d}x\,\ln_\phi(x),
\label {entdef}
\end{eqnarray}
Let us shortly analyse this expression.
Introduce a function $\xi$ given by
\begin{eqnarray}
\frac 1{\xi(u)}
=\int_0^1{\rm d}x\,\frac x{\phi(xu)}
=\frac 1u\left(\ln_\phi(u)-\frac 1u\int_0^u{\rm d}w\,\ln_\phi(w)\right)
\end{eqnarray}
Note that $\xi(u)$ is a positive increasing function of $u$.
One has
\begin{eqnarray}
I_\phi(\rho)
&=&-\Tr\rho \ln_\xi(\rho).
\label {altent}
\end{eqnarray}
This expression looks familiar.
In case of the natural logarithm is $\phi(x)=x$.
This implies that also $\xi(x)=x$. Then (\ref {entdef})
reduces to the conventional definition. From (\ref {altent}) follows also that
$I_\phi(\rho)\ge 0$ because $\ln_\xi(\rho)\le 0$.

Relative entropy, also called divergence, is defined by
\begin{eqnarray}
I_\phi(\rho||\sigma)&=&I_\phi(\sigma)-I_\phi(\rho)
-\Tr(\rho-\sigma)\ln_\phi(\sigma).
\label {relent}
\end{eqnarray}
Its basic property is the following.

\begin {theorem}{Proposition} For any pair of density operators $\rho$ and
$\sigma$ is
\begin{eqnarray}
I_\phi(\rho||\sigma)\ge 0
\end{eqnarray}
\end {theorem}

\begin {proof}
From
\begin{eqnarray}
\frac {{\rm d}\,}{{\rm d}u}u\ln_\xi u
=\int_0^1{\rm d}x\,x\int_1^{u}{\rm d}v\,
\left(\frac 1{\phi(xv)}-\frac 1{\phi(xu)}\right)
\end{eqnarray}
follows that the function $f(u)=u\ln_\xi(u)$ is convex.
Hence Klein's inequality (see e.g. \cite {RD69}, 2.5.2)
\begin {eqnarray}
\Tr \bigg(f(\rho)-f(\sigma)-(\rho-\sigma)f'(\sigma)\bigg)\ge 0
\end {eqnarray}
implies
\begin{eqnarray}
-I_\phi(\rho)+I_\phi(\sigma)-\Tr(\rho-\sigma)\left(\ln_\xi\sigma
+\frac {\sigma}{\xi(\sigma)}\right)\ge 0.
\end{eqnarray}
The proof follows now from the identity
\begin{eqnarray}
\ln_\phi u=\ln_\xi u+\frac u{\xi(u)}-\frac 1{\xi(1)}.
\end{eqnarray}
\end {proof}

This result can be used to prove concavity of the entropy functional
$I_\phi(\rho)$.

\begin {theorem}{Corollary}
For any $\lambda$ satisfying $0\le \lambda\le 1$ is
\begin{eqnarray}
I_\phi(\lambda\sigma+(1-\lambda)\rho)
\ge
\lambda I_\phi(\sigma)+(1-\lambda)I_\phi(\rho).
\label {concent}
\end{eqnarray}
\end {theorem}

\begin {proof}
Let $\tau=\lambda\sigma+(1-\lambda)\rho$. Applying twice the proposition,
in combination with definition (\ref {relent}), one obtains
\begin{eqnarray}
I_\phi(\tau)-I_\phi(\rho)&\ge&\Tr(\rho-\tau)\ln_\phi(\tau)\cr
I_\phi(\tau)-I_\phi(\sigma)&\ge&\Tr(\sigma-\tau)\ln_\phi(\tau).
\end{eqnarray}
By taking a convex combination of these two expressions (\ref {concent})
follows.
\end {proof}

Finally, note that a short calculation shows that
\begin{eqnarray}
I_\phi\left(\rho+u^k\frac{\partial\rho}{\partial\theta^k}\bigg|\bigg|\rho\right)
+
I_\phi\left(\rho\bigg|\bigg|\rho+u^k\frac{\partial\rho}{\partial\theta^k}\right)
&=&\frac 1{Z(\theta)}u^ku^lg_{kl}(\theta)+{\rm o}(u^2).
\end{eqnarray}
This expression links relative entropy to generalized Fisher information.

\section {Example}

The obvious example is obtained by taking $\phi(u)=u^q$, with $0<q<2$ ($q=1$
corresponds with the standard case). In this case the notations $\ln_q$ and
$\exp_q$ are used instead of $\ln_\phi$ resp.~$\exp_\phi$.
A short calculation gives
\begin{eqnarray}
\ln_q(u)=\frac 1{1-q}\left(u^{1-q}-1\right).
\end{eqnarray}
This particular definition of deformed logarithm has been introduced in
\cite {TC94}.
An easy integration gives $\xi(u)=(2-q)u^q$. Hence, in this case the functions
$\phi$ and $\xi$ differ only by a constant factor.
The resulting entropy functional is
\begin{eqnarray}
I_\phi(\rho)=-\frac 1{2-q}\Tr\rho\ln_q(\rho)
=\frac 1{2-q}\frac 1{q-1}\left(\Tr\rho^{2-q}-1\right).
\end{eqnarray}
This is, up to some changes in notation and a proportionality factor $1/(q-1)$,
the entropy studied in the context of Tsallis' thermostatistics \cite {TC88}.
The corresponding expression for relative entropy is
\begin{eqnarray}
I_\phi(\rho||\sigma)
=\frac 1{(q-1)(2-q)}
\Tr\left((q-1)\sigma^{2-q}-\rho^{2-q}+(2-q)\rho\sigma^{1-q}\right).
\end{eqnarray}

\section {Variational principle and duality}

Let us now show that the $\phi$-exponential family satisfies a maximum
entropy principle.

\begin {theorem}{Proposition}
Let $(\rho_\theta)_\theta$ be a $\phi$-exponential family of density operators
with two-by-two commuting Hamiltonians $H_1,\cdots,H_N$, and with
averages $\langle H_k\rangle_\theta$ equal to the gradients of the potential
$F(\theta)$.
There exists a constant $F_0$ such that
\begin{eqnarray}
F(\theta)=F_0+\min_\rho\{\Tr\rho\theta^k H_k-I_\phi(\rho)\}.
\end{eqnarray}
The minimum is attained for $\rho=\rho_\theta$. In particular, $F(\theta)$
is a convex function of $\theta$ and $\rho=\rho_\theta$ maximizes $I_\phi(\rho)$
under the constraint that $\Tr\theta^k\rho H_k=\Tr\theta^k\rho_\theta H_k$.
\end {theorem}

\begin {proof}
From $I(\rho||\rho_\theta)\ge 0$ follows
\begin{eqnarray}
0&\le&I_\phi(\rho_\theta)-I_\phi(\rho)
-\Tr(\rho-\rho_\theta)\ln_\phi(\rho_\theta).
\end{eqnarray}
But one has
\begin{eqnarray}
\ln_\phi(\rho_\theta)=G(\theta)-\theta^kH_k.
\end{eqnarray}
Hence one obtains
\begin{eqnarray}
0&\le&I_\phi(\rho_\theta)-I_\phi(\rho)
+\Tr(\rho-\rho_\theta)\theta^kH_k.
\end{eqnarray}
This proves that for all $\rho$
\begin{eqnarray}
\Tr\rho_\theta\theta^kH_k-I_\phi(\rho_\theta)
\le \Tr\rho\theta^kH_k-I_\phi(\rho).
\end{eqnarray}

Next note that
\begin{eqnarray}
\frac {\partial F}{\partial\theta^k}
=\langle H_k\rangle_\theta
=\frac {\partial\,}{\partial\theta^k}
\bigg(\theta^l\langle H_l\rangle_\theta-I_\phi(\rho_\theta)\bigg)
\label {temptheo}
\end{eqnarray}
because one has
\begin{eqnarray}
\frac {\partial \,}{\partial\theta^k}I_\phi(\rho_\theta)
&=&-\frac {\partial \,}{\partial\theta^k}\Tr\rho_\theta\ln_\xi\rho_\theta\cr
&=&-\Tr\left(\ln_\xi \rho_\theta+\frac 1{\xi(\rho_\theta)}\rho_\theta\right)
\frac {\partial \,}{\partial\theta^k}\rho_\theta\cr
&=&-\Tr\left(\ln_\phi \rho_\theta\right)
\frac {\partial \,}{\partial\theta^k}\rho_\theta\cr
&=&-\Tr\left(G(\theta)-\theta^lH_l\right)
\frac {\partial \,}{\partial\theta^k}\rho_\theta\cr
&=&\theta^l\frac {\partial \,}{\partial\theta^k}\langle H_l\rangle_\theta.
\label {ider}
\end{eqnarray}
From (\ref {temptheo}) follows that there exists a constant $F_0$ for which
\begin{eqnarray}
F(\theta)=F_0+\theta^l\langle H_l\rangle_\theta-I_\phi(\rho_\theta)
\label {theof}
\end{eqnarray}
holds for all $\theta$. This finishes the proof.

\end {proof}

Introduce dual coordinates $\eta_k$ (dual to $\theta_k$) by (see (\ref {aven}))
\begin{eqnarray}
\eta_k=\langle H_k\rangle_\theta=\frac {\partial F}{\partial\theta^k}.
\label {etadef}
\end{eqnarray}
They satisfy (use (\ref {expg}, \ref {i2}))
\begin{eqnarray}
\frac {\partial\eta_l}{\partial\theta^k}
=\frac{\partial^2 F}{\partial\theta^k\partial\theta^l}
=Z(\theta)\bigg([H_k]_\theta[H_l]_\theta-[H_kH_l]_\theta\bigg)
=-\frac 1{Z(\theta)}g_{kl}(\theta).
\label {ortho}
\end{eqnarray}
These are the orthogonality relations between the two sets of coordinates.
The dual of relation (\ref {etadef}) is
\begin{eqnarray}
\theta^k=\frac {\partial\,}{\partial\eta_k}I_\phi(\rho_\theta).
\label {dualrel}
\end{eqnarray}
Indeed, using (\ref {ider}, \ref {ortho}) one obtains
\begin {eqnarray}
\frac {\partial\,}{\partial\eta_k}I_\phi(\rho_\theta)
=\frac {\partial\theta^l}{\partial\eta_k}
\frac {\partial \,}{\partial\theta^l}I_\phi(\rho_\theta)
=-Z(\theta)g^{kl}\frac {\partial^2 F}{\partial\theta^l\partial\theta^m}\theta^m,
\end {eqnarray}
where $g^{kl}$ is the inverse matrix of $g_{kl}$.
Next use (\ref {expg}) to obtain (\ref {dualrel}).

Let $E(\eta)=I_\phi(\rho_\theta)$. Then one can write (\ref {theof}) as
(putting $F_0=0$)
\begin{eqnarray}
F(\theta)+E(\eta)=\theta^k\eta_k.
\end{eqnarray}
This shows that $F(\theta)$ and $E(\eta)$ are each others Legendre transform.
This generalizes the well-known result of standard thermostatistics that
free energy is the Legendre transform of equilibrium entropy.

\section{Summary and discussion}

This paper introduces the notion of $\phi$-exponential family of
density operators $\rho_\theta$. It depends on an arbitrary non-decreasing
non-negative
function $\phi$ of the positive reals and on a set of Hamiltonians $H_k$.
The analogy with the standard exponential family is enhanced by using the
notion of deformed exponential function $\exp_\phi$.
The definition is motivated
by showing that the $\phi$-exponential family, together with a family
of escort density operators, optimizes a generalized version of the
well-known lower bound of Cram\'er and Rao.

Generalized definitions of entropy (\ref {entdef}) and of relative entropy
(\ref {relent}) have been given. The definitions are such that the
$\phi$-exponential family of density operators satisfies a maximum entropy
principle and that the function $F(\theta)$ and entropy
$I_\phi(\rho_\theta)$ are each others Legendre transforms. The relation
between $F(\theta)$ and relative entropy is
\begin{eqnarray}
\theta^k\Tr\rho H_k-I_\phi(\rho)=F(\theta)+I_\phi(\rho||\rho_\theta).
\end{eqnarray}
The latter has the interpretation that free energy is
minimal in equilibrium (i.e., when $\rho=\rho_\theta$).

The assumption has been made that the Hamiltonians $H_k$ are two-by-two
commuting. As a consequence, the quantum information 'manifold'
$(\rho_\theta)_\theta$ is still abelian. This is clearly too restrictive
for a fully quantum-mechanical theory.
Further work is needed to remove this restriction.



\end {document}